\documentclass{article}

\usepackage[nonatbib]{arxiv}

\usepackage[utf8]{inputenc} 
\usepackage[T1]{fontenc}    
\usepackage{hyperref}       
\usepackage{url}            
\usepackage{booktabs}       
\usepackage{amsfonts}       
\usepackage{nicefrac}       
\usepackage{microtype}      
\usepackage{lipsum}
\usepackage{graphicx}
\usepackage{amsthm,amsmath}

\title{Feature Trajectory Dynamic Time Warping for Clustering of Speech Segments}

\author{
  Lerato Lerato, Thomas ~Niesler \\
  Department of Electrical and Electronic Engineering\\
  University of Stellenbosch\\
  Stellenbosch,  South Africa \\
  \texttt{llerato@sun.ac.ac.za},  \texttt{trn@sun.ac.za} \\
}

\begin{document}
\maketitle

\begin{abstract} 
Dynamic time warping (DTW) can be used to compute the similarity between two sequences of generally differing length.
We propose a modification to DTW that performs individual and independent pairwise alignment of feature trajectories.
The modified technique, termed feature trajectory dynamic time warping (FTDTW), is applied as a similarity measure in the agglomerative hierarchical clustering of speech segments.
Experiments using MFCC and PLP parametrisations extracted from TIMIT and from the Spoken Arabic Digit Dataset (SADD) show consistent and statistically significant improvements in the quality of the resulting clusters in terms of F-measure and normalised mutual information (NMI).
\end{abstract}

\section{Introduction}
\label{sec1}

Dynamic Time Warping (DTW) is a method of optimally aligning two distinct time series of generally different length. 
In addition to the alignment, DTW computes a score indicating the similarity of the two sequences. 
This ability to quantify the similarity between time series has led to the application of DTW in automatic speech recognition (ASR) systems several decades ago \cite{sakoe1978dynamic,DTW_speech_Myers}. 
It has remained popular in this field, with more recent developments reported in \cite{muda2010voice} and \cite{zhang2014one}.

DTW has also found application in fields related to ASR. 
For example, it has been used successfully in keyword spotting and information retrieval (IR) systems \cite{zhang2009unsupervised,anguera2013information,lee2015spoken}. 
To accomplish IR, sub-sequences in a speech signal that match a template with certain degree of time warping are detected. 

In the related task of acoustic pattern discovery, DTW can be allowed to consider multiple local alignments between speech signals during the overall search \cite{Park_pattern_discover}. 
In this way DTW can find similar segment pairs in speech audio, followed by a clustering step \cite{walter2013hierarchical}. 
The resulting cluster labels are used to train hidden Markov models (HMMs).

In an effort to improve performance, several variations of DTW have been proposed since its inception. 
For example, a one-against-all index (OAI) for each time series under consideration is proposed in \cite{zhang2014one}.
The OAI is subsequently used to weight the corresponding DTW alignment score in a speech recognition system. 
DTW has also been modified to allow the direct matching of points along the best alignment for use in a signature verification system \cite{shanker2007off}.
A stability function is subsequently applied, and the resulting score is used as a similarity measure. 

We describe a modification of DTW and demonstrate its improved performance when used as a similarity measure to cluster speech segments. 
Our DTW modification exploits the asynchronous temporal structure of features extracted from speech. 
Related work has considered such feature trajectories by training separate hidden Markov models (HMMs) for each MFCC feature dimension \cite{Sagayama99Asynchron}. 
This work reports improvements in both phoneme and word recognition.
The clustering of speech segments also has several useful applications in ASR \cite{SvendsenICASSP1989,PaliwalASWU1990,Wang2013Posteriorgrams}.
Recently it has been particularly useful in the automatic discovery of sub-word units \cite{LivescusummaryofMyWork,Kamper2014clustering}. 


Section \ref{sec_cdtw} reviews the standard formulation of DTW and Section \ref{sec_ftdtw} describes our proposed modification. 
Section \ref{sec_eval} presents the evaluation tools we employ and Section \ref{sec_data} describes the data we use for experimentation.
Section \ref{sec_experiments} presents an experimental evaluation of the proposed method. 
Section \ref{sec_conclusion} discusses the results and concludes the paper.

\section{Classical Dynamic Time Warping (DTW)}
\label{sec_cdtw}

We consider speech segments as temporal sequences of multidimensional feature vectors in the Euclidean space. 
Sequences are of arbitrary and generally different length, but all vectors are of equal dimension. 
The DTW algorithm recursively determines the best alignment between two such vector time series by minimizing a cumulative path cost that is commonly based on Euclidean distances between time aligned vectors \cite{DTW_speech_Myers,keogh2001derivative}. 

Consider $N$ such sequences $\mathbf{X}_i$, $i=1,2,...,N$, each composed of $T_i$  feature vectors, as defined in Equation \ref{eq:FeatureSet}.

\vspace*{-2mm}
\begin{equation}
\label{eq:FeatureSet}
	\mathbf{X}_i =\lbrace \mathbf{x}_{i1},\mathbf{x}_{i2},...,\mathbf{x}_{iT_i} \rbrace, \ \ \ i=1,2,...,N
\end{equation}

Each feature vector $\mathbf{x}_{it}$ has \emph{m} dimensions, as indicated in Equation \ref{eq:FeatureVec}.

\vspace*{-2mm}
\begin{equation}
\label{eq:FeatureVec}
	\mathbf{x}_{it} = \Big \langle x_{it}^{(1)},x_{it}^{(2)},...,x_{it}^{(m)} \Big \rangle,\ \ \ t=1,2,..,T
\end{equation} 
\vspace*{-1mm}

Two sequences $\mathbf{X}_i$ and $\mathbf{X}_j$ are aligned by constructing a $T_i$-by-$T_j$ distance matrix $D_{ij}(p,q)$ whose entries contain the distances $d(\mathbf{x}_{ip}, \mathbf{x}_{jq})$. 
Typical choices for $d$ are the Euclidean distance and the Manhattan distance. 
A matrix of minimum accumulated distances $\gamma_{ij}(p,q)$ is then constructed by considering all paths from $D_{ij}(1,1)$ to $D_{ij}(p,q)$. 
Using local and global path constraints, $\gamma_{ij}(p,q)$ is computed recursively according to the principle of dynamic programming, as shown in Equation \ref{eq:classicDTW} \cite{DTW_speech_Myers}.

\vspace*{-3mm}
\begin{equation}
\label{eq:classicDTW}
	\gamma_{ij}(p,q) = D_{ij}(\mathbf{x}_{ip}, \mathbf{x}_{iq}) \,+ \\ \min \big \lbrace \gamma_{ij}(p-1,q-1), \gamma_{ij}(p-1,q), \gamma_{ij}(p,q-1)\big \rbrace
\end{equation}
\vspace*{-1mm}

The similarity $DTW(\mathbf{X}_i,\mathbf{X}_j)$ between vector sequences $\mathbf{X}_i$ and $\mathbf{X}_j$ is then given by Equation \ref{eq:classicDTWscore}.
Here $K$ is the length of the optimal path from $D_{ij}(1,1)$ to $D_{ij}(T_i,T_j)$ and is used to normalise the similarity value.

\begin{equation}
\label{eq:classicDTWscore}
	DTW(\mathbf{X}_i,\mathbf{X}_j)=\frac{1}{K}\gamma_{ij} \big(T_i,T_j \big)  
\end{equation} 

\begin{figure}[!htb]
\centering
\includegraphics[scale=0.85]{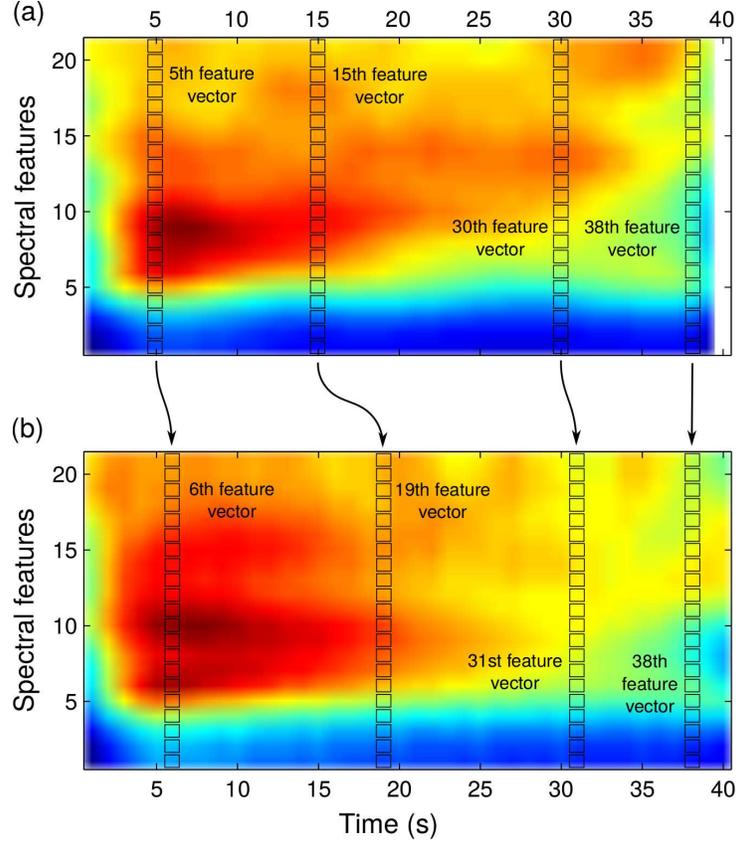}
\caption{Alignment by classical DTW of spectral features extracted from the triphone \emph{b-aa+dx} as uttered by (a) male speaker mrfk0 and (b) by female speaker fdml0 in the TIMIT corpus.}
\label{FIG_classical_dtw}
\end{figure}

This standard formulation of dynamic time warping will in the remainder of the paper be referred to as \textit{classical} DTW. 
Figure \ref{FIG_classical_dtw} shows the classical DTW alignment between two different sequences of 21-dimensional spectral feature vectors representing the same sound uttered by different speakers.
These spectral features are obtained by straightforward binning of the short-time power spectra.
To avoid clutter, the alignment of just four of the feature vectors is shown.

\section{Feature Trajectory DTW (FTDTW)}
\label{sec_ftdtw}

We define a feature trajectory $X_i^{(l)}$ as the time series obtained when considering the $l$-th element of each feature vector in a sequence $\mathbf{X}_i$, as shown in Equation \ref{eq:trajec}.
\begin{equation}
\label{eq:trajec}
	X_i^{(l)} =\Big \lbrace x_{i1}^{(l)}, x_{i2}^{(l)},...,x_{iT_i}^{(l)} \Big \rbrace, \ \ \ l = 1,2,...,m
\end{equation}

\begin{figure}[!htb]
\centering
\includegraphics[scale=0.85]{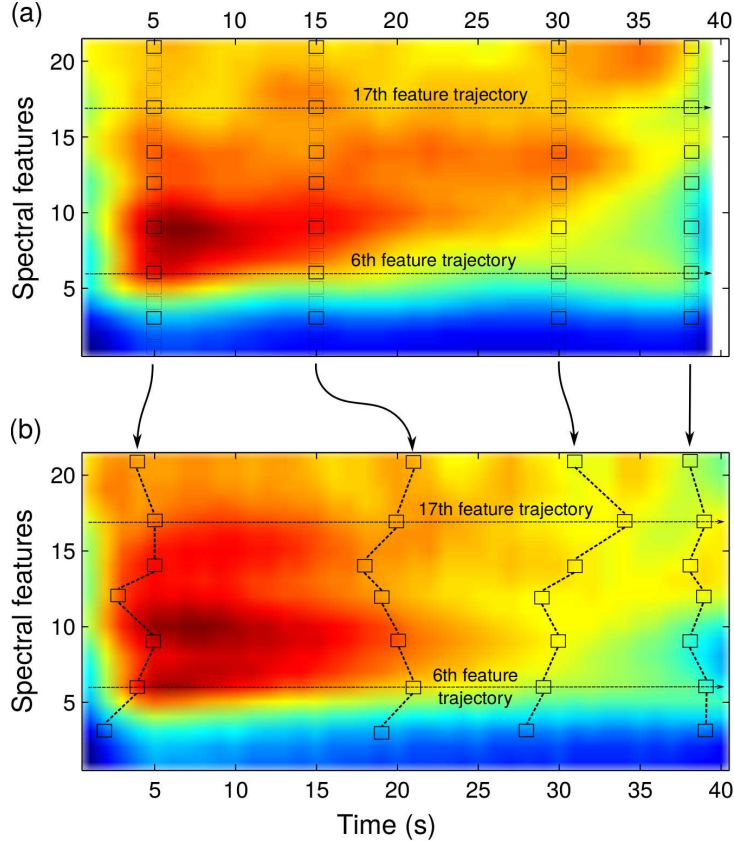}
\caption{Alignment by FTDTW of spectral features extracted from the triphone \emph{b-aa+dx} as uttered by (a) male speaker mrfk0 and (b) by female speaker fdml0 in the TIMIT corpus.}
\label{FIG_ftdtw}
\end{figure}

Hence $X_i^{(l)}$ is a 1-dimensional time series for feature \emph{l}. 
We now calculate the similarity of two feature vector sequences by applying classical DTW to each corresponding pair of feature trajectories, and subsequently normalise the sum, as shown in Equation \ref{eq:newdtw}.

\begin{equation}
\label{eq:newdtw}
	FTDTW(\mathbf{X}_i, \mathbf{X}_j)=\frac{1}{\beta} \sum_{l=1}^{m} DTW \Big \lbrace X_i^{(l)},X_j^{(l)} \Big \rbrace
\end{equation}
 
where $\beta=\sqrt[]{\sum_{l=1}^m K_l^2}$, $K$ is the path length and $DTW(.)$ is non-normalised classical DTW. 
As illustration, we repeat the alignment of the two speech segments shown in Figure~\ref{FIG_classical_dtw} with FTDTW.
Figure~\ref{FIG_ftdtw} (a) identifies seven features from each of the four feature vectors shown in Figure~\ref{FIG_classical_dtw} (a). 
Figure~\ref{FIG_ftdtw} (b) demonstrates how each of these seven features align with the second speech segment.
The features themselves are the same as those illustrated in Figure \ref{FIG_classical_dtw}.  
For the illustrated example, application of Equation \ref{eq:newdtw} involves 21 separate alignments, each between corresponding features trajectories as also indicated in Figure \ref{FIG_ftdtw}. 
The resulting 21 scores are summed and normalised by $\beta$. 
Figure~\ref{FIG_ftdtw} illustrates how, in contrast to classical DTW, FTDTW does not require features coincident in time in one segment to align with features in the other segment also coincident in time.

\section{Evaluation}
\label{sec_eval}
We evaluate the effectiveness of our proposed modification to DTW by using it to compute similarities between speech segments, and then using these similarities to perform agglomerative hierarchical clustering \cite{JainBook1988,murtagh2011methods}. 
We will cluster speech segments corresponding to triphones extracted from the TIMIT corpus as well as isolated digits extracted from the Spoken Arabic Digit Dataset (SADD).
Since the phonetic alignment is provided in the former and the word alignments in the latter, the ground truth is available. 
Hence we can use the external metrics F-measure and normalised mutual information (NMI) to quantify the quality of the resulting clusters \cite{Amigo_Cluster_eval_compa,Fmeasure_Larsen,NMIVinhinformationtheoretic}.

\subsection{Agglomerative Hierarchical Clustering (AHC)}
In AHC, the agglomeration of data objects (speech segments in the case of our experimental evaluation) is initialised by the assumption that each object is the sole occupant of its own cluster. 
A binary tree referred to as a dendrogram is created by successively merging the closest cluster pairs until a single cluster remains \cite{xuClustering2005survey}. 
We use the popular Ward method to quantify inter-cluster similarity \cite{murtaghWard2014}. 
The input to the AHC algorithm is a symmetric $N \times N$ proximity matrix populated by the values of $DTW(\cdot,\cdot)$ or $FTDTW(\cdot,\cdot)$ and the output consists of the $R$ clusters.

\subsection{F-measure}
The F-measure is based on the quantities precision (PR) and recall (RE). Precision indicates the degree to which a cluster is dominated by a particular class, while recall indicates the degree to which a particular class is concentrated in a specific cluster. Precision and recall  are defined in Equations \ref{eq:precision} and \ref{eq:recall} respectively.

\vspace*{-2mm}
\begin{equation}
\label{eq:precision}
	PR(r,v)=\frac{n_{rv}}{n_r}
\end{equation}

\begin{equation}
\label{eq:recall}
	RE(r,v)=\frac{n_{rv}}{n_v}
\end{equation}

Here $n_{rv}$ indicates the number of objects of class $v$ in cluster $r$; $n_r$ and $n_v$ the number of objects in cluster $r$ and class $v$ respectively. The F-measure ($F$) is given in Equation \ref{eq:Fmeasure}.
\begin{equation}
\label{eq:Fmeasure}
	F(r,v)=\frac{2 \times RE(r,v) \times PR(r,v)}{ RE(r,v)+PR(r,v)}
\end{equation}
When the clusters are perfect, $n_{rv}=n_r=n_v$ and hence $F(r,v)=1$.

\subsection{Normalised mutual information}
Normalised mutual information (NMI) employs the following formulations:
\begin{itemize}
\item the set of $R$ clusters $\mathbf{G}=\lbrace G_1,G_2,...,G_R \rbrace$, and
\item the set of $V$ classes $\mathbf{C}=\lbrace C_1,C_2,...,C_V \rbrace$  representing ground truth.
\end{itemize}
NMI is based on the mutual information, $I(\mathbf{G},\mathbf{C})$ between classes and clusters  \cite{NMIVinhinformationtheoretic,IRBook}. The mutual information is not sensitive to varying  number of clusters and therefore it is normalised by a factor based on the cluster entropy $H(\mathbf{G})$ and  class entropy $H(\mathbf{C})$. These entropies measure cluster and class cohesiveness respectively. The NMI criterion is given in Equation \ref{eq:NMI}.

\begin{equation}
\label{eq:NMI}
	NMI(\mathbf{G},\mathbf{C})=\frac{2I(\mathbf{G},\mathbf{C})}{\left[ H(\mathbf{G}) + H(\mathbf{C})\right]}
\end{equation}

The mutual information $I(\mathbf{G},\mathbf{C})$ and the entropies $H(\mathbf{G})$ and $H(\mathbf{C})$  are given in Equations \ref{eq:Info}, \ref{eq:Entropy} and \ref{eq:Entropy2} respectively.
\begin{equation}
\label{eq:Info}
	I(\mathbf{G},\mathbf{C})=\sum_{r \in \mathbf{G}} \sum_{v \in \mathbf{C}} P(G_r)P(C_v) \log \frac{P(G_r \cap C_v)}{P(G_r)P(C_v)}
\end{equation}

In Equation \ref{eq:Info}, $P(G_r)$, $P(C_v)$ and  $P(G_r \cap C_v)$ are the probabilities of a segment belonging to cluster $G_r$, class $C_v$ and the intersection of $G_r$ and $C_v$ respectively.

\begin{equation}
\label{eq:Entropy}
	H(\mathbf{G})= -\sum_{r \in \mathbf{G}}P(G_r) \log P(G_r)
\end{equation}

\begin{equation}
\label{eq:Entropy2}
	H(\mathbf{C})= -\sum_{v \in \mathbf{C}}P(C_v) \log P(C_v)
\end{equation}

It can be shown that $I(\mathbf{G},\mathbf{C})$ is zero when the clustering is random with respect to class membership and that it achieves a maximum of $1.0$ for perfect clustering \cite{IRBook}.

\section{Data}
\label{sec_data}
Our first set of experiments uses speech segments taken from the TIMIT speech corpus \cite{garofolo1993darpa}. 
TIMIT has been chosen because it includes accurate time-aligned phonetic transcriptions, meaning that both phonetic labels and their start/end times are known. 
As our desired clusters we use triphones, which are phones in specific left and right contexts \cite{triphone_clust_Imperl}.
We consider triphones that occur at least 20 times and at most 25 times in the corpus. 
This leads to an evenly balanced set of 8772 speech segments, which also corresponds approximately to the number of segments in our second set of experiments.  

For comparison and confirmation purposes, we performed a second set of experiments using the Spoken Arabic Digit Dataset (SADD) \cite{Lichman2013}. 
SADD consists of 8800 utterances already parametrised as 13-dimensional MFCCs. 
The utterances were spoken by 44 male and 44 female Arabic speakers. 
Each utterance in the SADD corresponds to a single Arabic digit, and will therefore be considered to be a single segment in our experiments. 
Each digit (0 to 9) was uttered ten times by each speaker. 

A third set of experiments is based on 10 independent subsets of speech segments drawn from the TIMIT SI and SX utterances, irrespective of occurrence frequency.
This better represents the unbalanced distribution of triphones that may be expected in unconstrained speech.
Table \ref{tab_data} summarises the datasets used in each of the three sets of experiments.

\begin{table}[!h]
\centering
\caption{Datasets used for experimental evaluation.}
\label{tab_data}
\begin{tabular}{ll}
\hline
Dataset & Description  \\
\hline
1 & 8772 TIMIT triphones (evenly balanced).\\
2 & 8800 SADD isolated digits (evenly balanced). \\
3 & 123182 TIMIT SI and SX triphones divided randomly into 10 subsets (not evenly balanced).\\
\hline
\end{tabular}
\end{table}
 
We considered two feature vector parametrisations popular in the field of speech processing, namely mel frequency cepstral coefficients (MFCCs) and perceptual linear  prediction (PLP) coefficients \cite{davis1980comparison,hermansky1990perceptual}. 
For the former, log frame energy was appended to the first 12 MFCCs to produce a 13-dimensional feature vector. 
First and second differentials (velocity and acceleration) were subsequently added to produce the final 39-dimensional MFCC feature vector. 
For the latter, 13 PLP coefficients were considered, to which velocity and acceleration were added, again resulting in a 39-dimensional feature vector. 
One such feature vector was extracted for each 10ms frame of speech, where consecutive frames overlapped by 5ms. 
All TIMIT feature vectors were computed using HTK \cite{htkbook}. 
SADD provides pre-computed MFCC features, and hence PLP features were not used in the associated experiments.

\section{Experiments}
\label{sec_experiments}

To evaluate the performance of Feature Trajectory DTW (FTDTW) as an alternative to classical DTW as a similarity measure, we will employ it to perform AHC of the speech segments described in Section~\ref{sec_data}.
The quality of the automatically-determined clusters will be determined using the F-measure and in several cases also NMI.

\begin{figure}[!htb]
\centering
\resizebox{100mm}{!}{\includegraphics{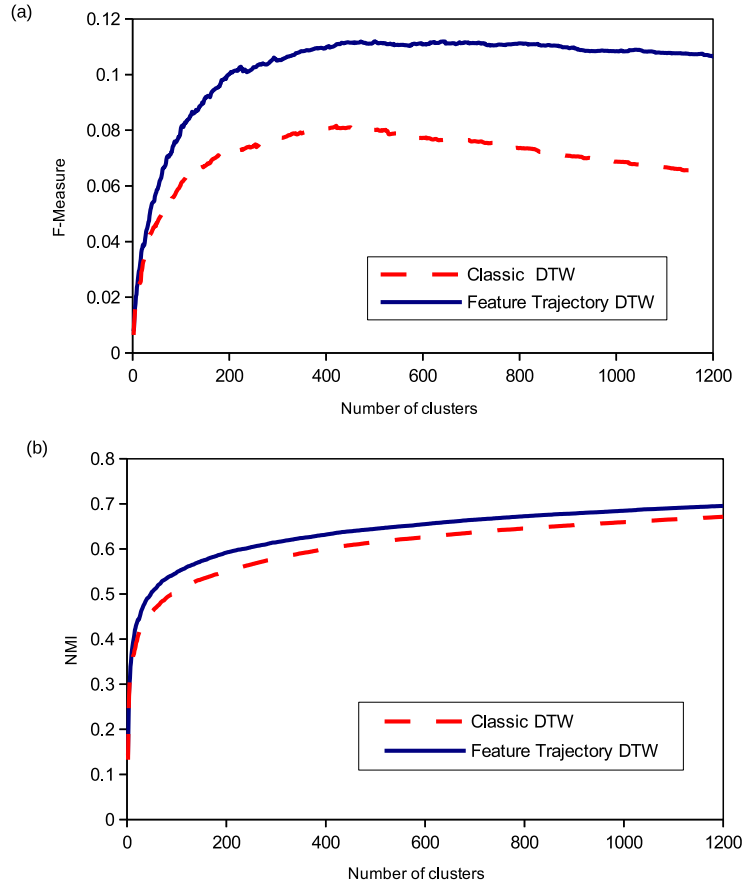}}
\caption{Clustering performance for Dataset 1 when using MFCC features in terms of (a) F-measure and (b) NMI.}
\label{Fig4}
\end{figure}

In a first set of experiments, we cluster Dataset~1 (Table~\ref{tab_data}). 
Figure \ref{Fig4} reflects the clustering performance in terms of (a) the F-measure  and (b) NMI, when using MFCCs as features.  
Both the F-measure and NMI are plotted as a function of the number of clusters.
Note that the F-measure continues to decline as the number of clusters exceeds 1200.
 

Figures \ref{Fig4}(a) and \ref{Fig4}(b) show that FTDTW improves on the performance of classical DTW in this clustering task in terms of both F-measure and NMI.  Especially in terms of F-measure, this improvement is substantial.

\begin{figure}[!htb]
\centering
\resizebox{100mm}{!}{\includegraphics{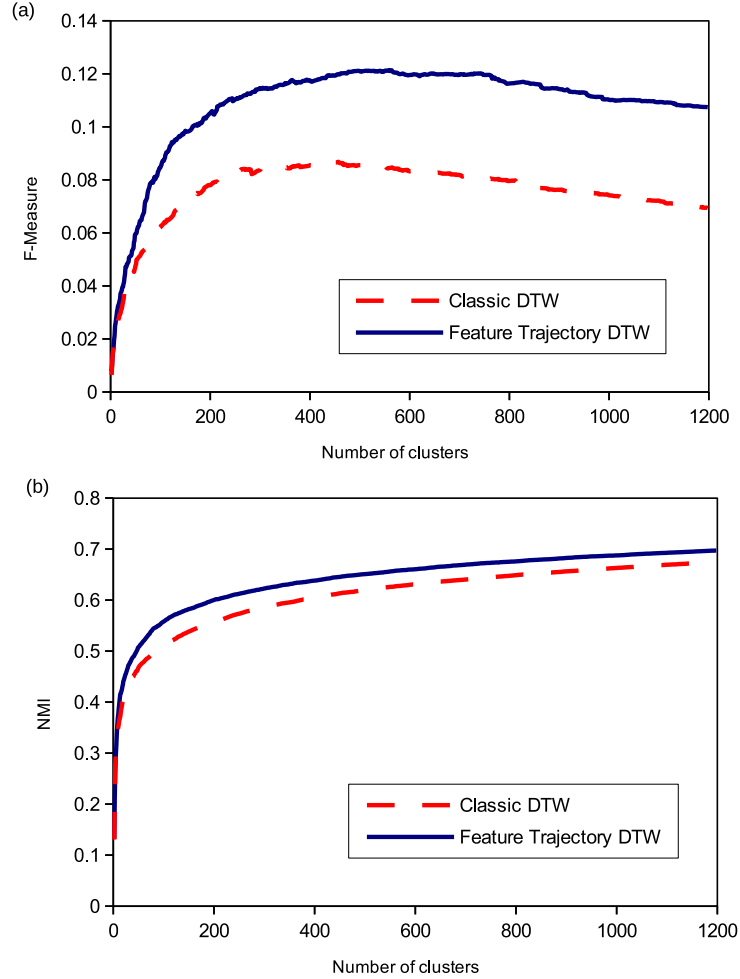}}
\caption{Clustering performance for Dataset 1 when using PLP features in terms of (a) F-measure and (b) NMI.}
\label{Fig5}
\end{figure}

A corresponding set of experiments using PLP features was carried out for Dataset~1, and the results are shown in Figure \ref{Fig5}. 
The same trends seen for MFCCs in Figure \ref{Fig4} are observed, with substantial improvements particularly in terms of F-measure.  

\begin{figure}[!htb]
\centering
\resizebox{100mm}{!}{\includegraphics{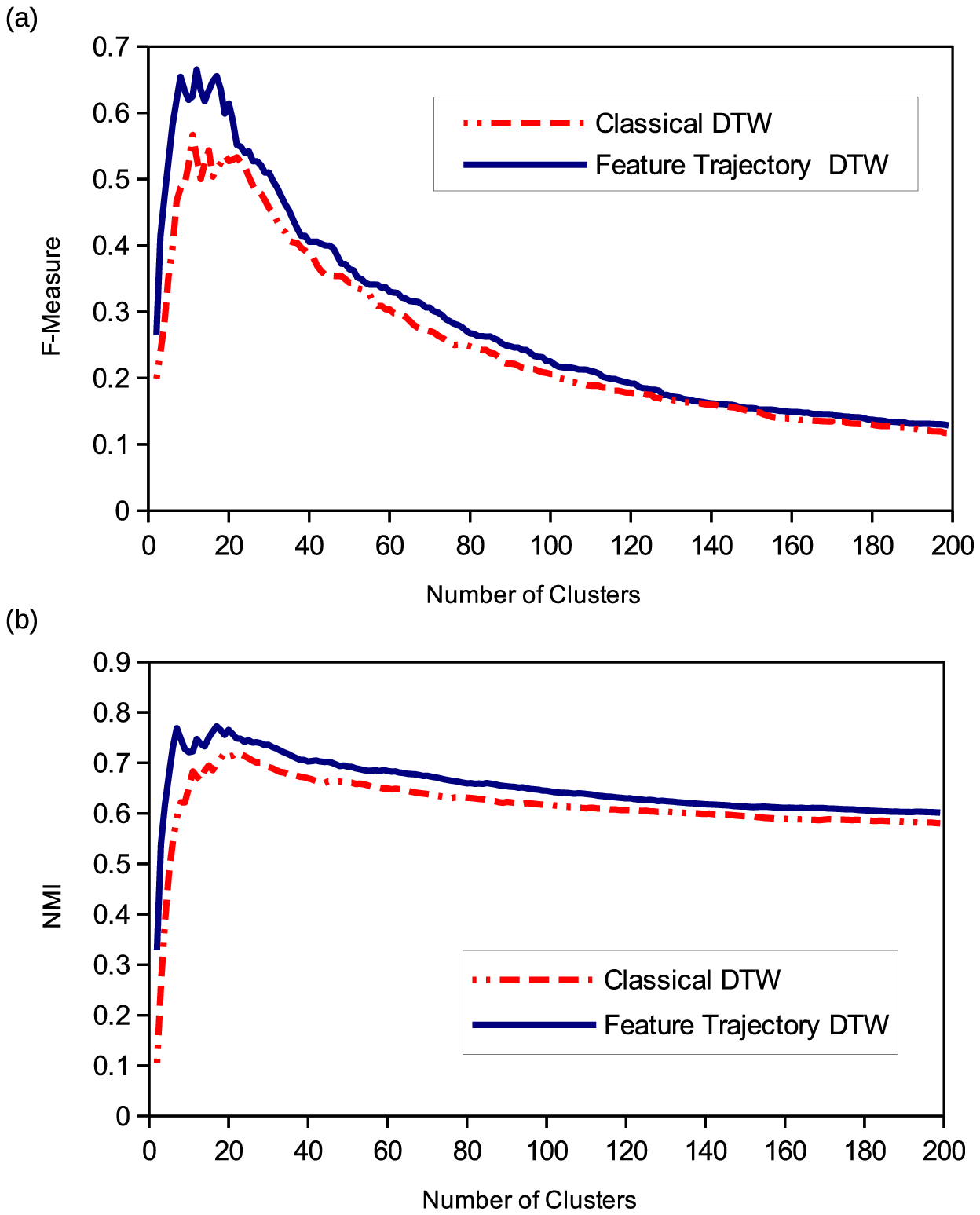}}
\caption{Clustering performance for Dataset 2 in terms of (a) F-measure and (b) NMI.}
\label{Fig6}
\end{figure}

In a second set of experiments, we clustered Dataset~2 (Table~\ref{tab_data}) which consists of isolated Arabic digits.
Figure \ref{Fig6} indicates the clustering performance, both in terms of  F-measure and NMI for this dataset. 
Again we observe that FTDTW outperforms classical DTW in terms of both F-measure and NMI in practically all cases.  

\begin{figure}[!htb]
\centering
\resizebox{100mm}{!}{\includegraphics{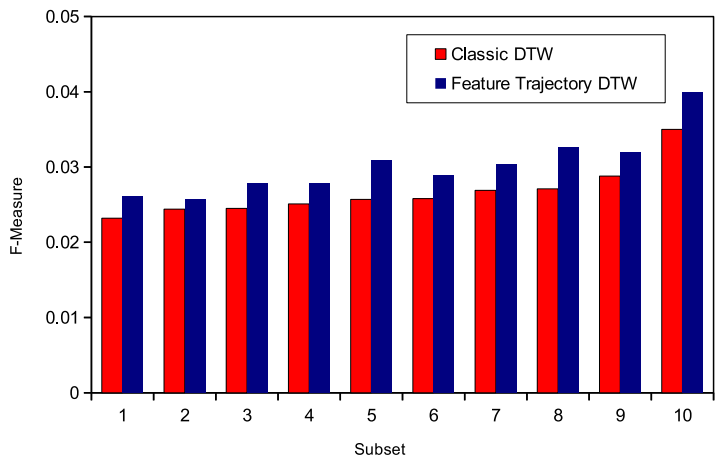}}
\caption{Clustering performance for the 10 independent subsets of Dataset 3 in terms of F-measure.}
\label{Fig7}
\end{figure}

In a third and final set of experiments, we considered Dataset~3 (Table~\ref{tab_data}).
The 10 independent subsets of the TIMIT training set each contained between 12034 and 12495 triphone segments. 
In contrast to the experiments for Dataset~1, all triphone tokens were considered irrespective of occurrence frequency.  
Furthermore, the number of clusters was chosen to be 2394, a figure which  corresponds to the number of triphone types with more than 10 occurrences in the data.
A single number of clusters, rather than a range as presented in Figures~\ref{Fig4},~\ref{Fig5} and~\ref{Fig6}, has been used here in order to make the required computations practical. 
Figure \ref{Fig7} presents the clustering performance for each of the 10 subsets in terms of  F-measure. 
We observe that FTDTW achieves an improvement over classical DTW in all cases.
A paired t-test indicated $p < 0.0001$, and hence the improvements are statistically highly significant. 
Similar improvements were observed in terms of NMI.

\section{Discussion and conclusions}
\label{sec_conclusion}

The experiments in Section~\ref{sec_experiments} have applied our modified DTW algorithm (FTDTW) to the clustering of speech segments. 
Our experiments show consistent and statistically significant improvement over the classical DTW baseline for both MFCC and PLP parametrisations and across three data sets.
We conclude that FTDTW is more effective as a similarity measure for speech signals than classical DTW.

Because classical DTW operates on a feature-vector by feature-vector basis, it enforces absolute temporal synchrony between the feature trajectories.
In contrast, FTDTW does not impose this synchrony constraint, but aligns feature trajectories independently on a pair-by-pair basis. 
Since FTDTW is observed to lead to better clusters in our experiments, we conclude that the strict temporal synchrony imposed by classical DTW is counter-productive in the case of speech signals.
We further speculate that segments of speech that human listeners would regard as similar also exhibit such differing time-scale warping among the feature trajectories.
It remains to be seen whether this decoupling of the feature trajectories is advantageous for signals other than speech.

Finally, and noting that it is not a focus of this paper, we may consider the maxima observed in the F-measure in Figure~\ref{Fig4} and \ref{Fig5}, and in both the F-measure and NMI in Figure~\ref{Fig6}.
A peak in the quality of the clusters as a function of the number of clusters may be taken to indicate the best estimate of the 'true' number of clusters in the data.
For the experiments using the MFCC parametrisation of Dataset~1 (Figure~\ref{Fig5}), we see that an optimum in the F-measure in reached at 501 and 421 clusters for FTDTW and classical DTW respectively. 
The 'true' number of clusters corresponds to the number of triphone types in Dataset~1, which is 404.
Hence both DTW formulations over-estimate the number of clusters.
A similar tendency is seen for the PLP parametrisations of the same dataset, where the F-measure peaks at 439 and 559 clusters for classical DTW and proposed DTW respectively, and also for Dataset~2 in Figure~\ref{Fig6}.

Although the ground truth is known, the class definitions (triphones for Datasets~1 and 3, and isolated digits for Dataset~2) may be called into question.
In particular, although all triphones correspond to acoustic segments from the same phone within the same left and right contexts, there are many other possible sources of systematic variability, such as the accent of the speaker.
Hence it may be reasonable to expect that a larger number of clusters is needed to optimally model the data.
To determine whether this is the case, the clusters should be used to determine acoustic models for an ASR system.
Then the performance of varying clusterings of the data can be compared by comparing the performance of the resulting ASR systems.
We intend to address this question in ongoing work.


\begin{thebibliography}{10}

\bibitem{sakoe1978dynamic}
Hiroaki Sakoe and Seibi Chiba.
\newblock Dynamic programming algorithm optimization for spoken word
  recognition.
\newblock {\em IEEE Transactions on Acoustics, Speech and Signal Processing},
  26(1):43--49, 1978.

\bibitem{DTW_speech_Myers}
C.~Myers, L.R. Rabiner, and A.E. Rosenberg.
\newblock Performance tradeoffs in dynamic time warping algorithms for isolated
  word recognition.
\newblock {\em IEEE Transactions on Acoustics, Speech, and Signal Processing},
  28(6):623--635, 1980.

\bibitem{muda2010voice}
Lindasalwa Muda, Mumtaj Begam, and I~Elamvazuthi.
\newblock Voice recognition algorithms using mel frequency cepstral coefficient
  ({MFCC}) and dynamic time warping ({DTW}) techniques.
\newblock {\em arXiv preprint arXiv:1003.4083}, 2010.

\bibitem{zhang2014one}
Xianglilan Zhang, Jiping Sun, and Zhigang Luo.
\newblock One-against-all weighted dynamic time warping for
  language-independent and speaker-dependent speech recognition in adverse
  conditions.
\newblock {\em PloS ONE}, 9(2):e85458, 2014.

\bibitem{zhang2009unsupervised}
Yaodong Zhang and James~R Glass.
\newblock Unsupervised spoken keyword spotting via segmental {DTW} on
  {G}aussian posteriorgrams.
\newblock In {\em Proc. IEEE Automatic Speech Recognition \& Understanding
  Workshop ({ASRU})}, pages 398--403, 2009.

\bibitem{anguera2013information}
Xavier Anguera.
\newblock Information retrieval-based dynamic time warping.
\newblock In {\em Proc. Interspeech}, pages 1--5, 2013.

\bibitem{lee2015spoken}
Lin-Shan Lee, James Glass, Hung-Yi Lee, and Chun-An Chan.
\newblock Spoken content retrieval—beyond cascading speech recognition with
  text retrieval.
\newblock {\em Audio, Speech, and Language Processing, IEEE/ACM Transactions
  on}, 23(9):1389--1420, 2015.

\bibitem{Park_pattern_discover}
A.~Park and J.~Glass.
\newblock Towards unsupervised pattern discovery in speech.
\newblock In {\em Proc. IEEE Workshop on Automatic Speech Recognition and
  Understanding ({ASRU})}, 2005.

\bibitem{walter2013hierarchical}
Oliver Walter, Timo Korthals, Reinhold Haeb-Umbach, and Bhiksha Raj.
\newblock A hierarchical system for word discovery exploiting {DTW}-based
  initialization.
\newblock In {\em Proc. IEEE Workshop on Automatic Speech Recognition and
  Understanding ({ASRU})}, pages 386--391, 2013.

\bibitem{shanker2007off}
A~Piyush Shanker and AN~Rajagopalan.
\newblock Off-line signature verification using {DTW}.
\newblock {\em Pattern Recognition Letters}, 28(12):1407--1414, 2007.

\bibitem{Sagayama99Asynchron}
Shigeki Sagayama, Shigeki Matsuda, Mitsuru Nakai, and Hiroshi Shimodaira.
\newblock Asynchronous-transition {HMM} for acoustic modeling.
\newblock In {\em International Workshop on Acoustic Speech Recognition and
  Understanding}, pages 99--102, 1999.

\bibitem{SvendsenICASSP1989}
T.~Svendsen, K.K. Paliwal, E.~Harborg, and P.O. Husoy.
\newblock An improved sub-word based speech recognizer.
\newblock In {\em Proc. IEEE International Conference on Acoustics, Speech and
  Signal Processing ({ICASSP})}, pages 729--732, 1989.

\bibitem{PaliwalASWU1990}
K.K. Paliwal.
\newblock Lexicon-building methods for an acoustic sub-word based speech
  recognizer.
\newblock In {\em Proc. IEEE International Conference on Acoustics, Speech and
  Signal Processing ({ICASSP})}, pages 108--111, 1990.

\bibitem{Wang2013Posteriorgrams}
H.~Wang, T.~Lee, C.~Leung, B.~Ma, and H.~Li.
\newblock Unsupervised mining of acoustic subword units with segment-level
  {G}aussian posteriograms.
\newblock In {\em Proc. Interspeech}, pages 2297--2301, 2013.

\bibitem{LivescusummaryofMyWork}
K.~Livescu, E.~Fosler-Lussier, and F.Metze.
\newblock Subword modeling for automatic speech recognition: Past, present, and
  emerging approaches.
\newblock {\em IEEE Signal Processing Magazine}, pages 44--57, 2012.

\bibitem{Kamper2014clustering}
Herman Kamper, Aren Jansen, Simon King, and Sharon Goldwater.
\newblock Unsupervised lexical clustering of speech segments using fixed
  dimensional acoustic embeddings.
\newblock In {\em Proc. IEEE Spoken Language Technology Workshop (SLT)}, 2014.

\bibitem{keogh2001derivative}
Eamonn~J Keogh and Michael~J Pazzani.
\newblock Derivative dynamic time warping.
\newblock In {\em SDM}, volume~1, pages 5--7. SIAM, 2001.

\bibitem{JainBook1988}
Anil~K. Jain and Richard~C. Dubes.
\newblock {\em Algorithms for Clustering Data}.
\newblock Prentice-Hall, Inc., Upper Saddle River, NJ, USA, 1988.

\bibitem{murtagh2011methods}
Fionn Murtagh and Pedro Contreras.
\newblock Methods of hierarchical clustering.
\newblock {\em arXiv preprint arXiv{:}1105.0121}, 2011.

\bibitem{Amigo_Cluster_eval_compa}
J.~Artiles E.~Amigo, J.~Gonzalo and F.~Verdejo.
\newblock A comparison of extrinsic clustering evaluation metrics based on
  formal constraints.
\newblock {\em Information Retrieval}, 12(4):461--486, 2009.

\bibitem{Fmeasure_Larsen}
B.~Larsen and C.~Aone.
\newblock Fast and effective text mining using linear-time document clustering.
\newblock In {\em Proc. ACM Conference on Knowledge Discovery and Data Mining
  (SIGKDD)}, pages 16--22, New York, USA, 1999.

\bibitem{NMIVinhinformationtheoretic}
N.~X.Vinh, J.~Epps, and J.~Bailey.
\newblock Information theoretic measures for clusterings comparison: Variants,
  properties, normalisation and correction for chance.
\newblock {\em Journal of Machine Learning Research}, 11:2837--2854, 2010.

\bibitem{xuClustering2005survey}
Rui Xu and Donald Wunsch.
\newblock Survey of clustering algorithms.
\newblock {\em IEEE Transactions on Neural Networks}, 16(3):645--678, 2005.

\bibitem{murtaghWard2014}
Fionn Murtagh and Pierre Legendre.
\newblock Ward’s hierarchical agglomerative clustering method: Which
  algorithms implement {W}ard’s criterion?
\newblock {\em Journal of Classification}, 31(3):274--295, 2014.

\bibitem{IRBook}
C.~D. Manning and P.~Raghavan.
\newblock {\em Introduction to Information Retrieval}.
\newblock Cambridge University Press, New York, USA, 2008.

\bibitem{garofolo1993darpa}
John~S Garofolo, Lori~F Lamel, William~M Fisher, Jonathon~G Fiscus, David~S
  Pallett, Nancy Dahlgren, and Victor Zue.
\newblock {DARPA} {TIMIT} acoustic-phonetic continous speech corpus {LDC93S1}.
\newblock 1993.

\bibitem{triphone_clust_Imperl}
B.~Imperl, Z.~Kacic, B.~Horvat, and A.~Zgank.
\newblock Clustering of triphones using phoneme similarity estimation for the
  definition of a multilingual set of triphones.
\newblock {\em Speech Communication}, 39(4):353--366, 2003.

\bibitem{Lichman2013}
M.~Lichman.
\newblock {UCI} machine learning repository, 2013.

\bibitem{davis1980comparison}
Steven~B Davis and Paul Mermelstein.
\newblock Comparison of parametric representations for monosyllabic word
  recognition in continuously spoken sentences.
\newblock {\em IEEE Transactions on Acoustics, Speech and Signal Processing},
  28(4):357--366, 1980.

\bibitem{hermansky1990perceptual}
Hynek Hermansky.
\newblock Perceptual linear predictive ({PLP}) analysis of speech.
\newblock {\em Journal of the Acoustical Society of America}, 87(4):1738--1752,
  1990.

\bibitem{htkbook}
S.~J. Young, G.~Evermann, M.~J.~F. Gales, T.~Hain, D.~Kershaw, G.~Moore,
  J.~Odell, D.~Ollason, D.~Povey, V.~Valtchev, and P.~C. Woodland.
\newblock {\em The {HTK} Book, version 3.4}.
\newblock Cambridge University Engineering Department, Cambridge, UK, 2006.

\end{thebibliography}

\end{document}